\input phyzzx.tex
\tolerance=1000
\voffset=-0.0cm
\hoffset=0.7cm
\sequentialequations

\def\t1{{\tilde 1}}

\def\t{\theta}

\REF{\CARL}{S. Carlip, Phys. Rev. Lett. {\bf 82} (1999) 2828, [arXiv:hep-th.9812013]; Class. Quant. Grav. {\bf 16} (1999) 3327,
[arXiv:gr-qc/9906126].}
\REF{\SOL}{S. Solodukhin, Phys. Lett. {\bf B454} (1999) 213, [arXiv:hep-th/9812056].}
\REF{\LEN}{L. Susskind, [arXiv:hep-th/9309145].}
\REF{\CFT}{E. Halyo, [arXiv:1502.01979]; [arXiv:1503.07808]; [arXiv:1506.05016].}
\REF{\SBH}{E. Halyo, A. Rajaraman and L. Susskind, Phys. Lett. {\bf B392} (1997) 319, [arXiv:hep-th/9605112].}
\REF{\HRS}{E. Halyo, B. Kol, A. Rajaraman and L. Susskind, Phys. Lett. {\bf B401} (1997) 15, [arXiv:hep-th/9609075].}
\REF{\EDI}{E. Halyo, Int. Journ. Mod. Phys. {\bf A14} (1999) 3831, [arXiv:hep-th/9610068]; Mod. Phys. Lett. {\bf A13} (1998), [arXiv:hep-th/9611175].}
\REF{\DES}{E. Halyo, [arXiv:hep-th/0107169].}
\REF{\UNI}{E. Halyo, JHEP {\bf 0112} (2001) 005, [arXiv:hep-th/0108167]; [arXiv:hep-th/0308166].}
\REF{\EDIH}{E. Halyo, [arXiv:1406.5763].}
\REF{\WAL}{R. M. Wald, Phys. Rev. {\bf D48} (1993) 3427, [arXiv:gr-gc/9307038]; V. Iyer and R. M. Wald, Phys. Rev. {\bf D50} (1994) 846, [arXiv:gr-qc/9403028]; Phys. Rev. {\bf D52} (1995) 4430, [arXiv:gr-qc/9503052].}
\REF{\EDIW}{E. Halyo. [arXiv:1403.2333].}
\REF{\CAR}{J. L. Cardy, Nucl. Phys. {\bf B463} (1986) 435.}
\REF{\FAT}{J. Maldacena and L. Susskind, Nucl. Phys. {\bf B475} (1996) 679, [arXiv:hep-th/9604042].}
\REF{\VER}{G. J. Turiaci and H. L. Verlinde, JHEP {\bf 1612} (2016) 110 [arXiv:1603.03020].}
\REF{\VIR}{A. Alekseev and S. L. Shatashvili, Comm, Math. Phys. {\bf 128} (1990) 197.}
\REF{\POL}{A. Almheiri and J. Polchinski, JHEP {\bf 11} (2015) 014, [arXiv:1402.6334]}
\REF{\MAL}{J. Maldacena, D. Stanford and Z. Yang, PTEP 2016 (2016) no.12, 12C104, [arXiv:1606.01857].}
\REF{\JT}{R. Jackiw, Nucl. Phys. {\bf B252} (1985) 343; C. Teitelboim, Phys. Lett. {B126} (1983) 41.}
\REF{\ADS}{E. Halyo, [arXiv:1706.07428]; [arXiv:1805.06079]; to appear.}
\REF{\CHA}{S. H. Shenker and D. Stanford, JHEP {\bf1403} (2014) 067, [arXiv:1306.0622]; JHEP {\bf 1412} (2014) 046,
[arXiv:1312.3296]; J. Maldacena, S. H. Shenker and D. Stanford, JHEP {\bf 1608} (2016) 106, [arXiv:1503.01409].}
\REF{\NEXT}{E. Halyo, in preparation.}
\REF{\SYK}{A. Kitaev, http://online.kitp.ucsb.edu/online/entangled15/kitaev/, http:// \hfill
online.kitp.ucsb.edu/online/entangled15/kitaev2/,
J. Madacena and D. Stanford, Phys.Rev. {\bf D94} (2016) no.10, 106002, [arXiv:1604.07818].} 
\REF{\LIO}{E. Halyo, [arXiv:1606.00792].}
\REF{\YMO}{L. D. Faddeev and A. Y. Volkov, Lett. Math. Phys. {\bf 32} (1994) 125, [arXiv:hep-th/9405087]; L. D. Faddeev, R. M. Kashaev andA. Y. Volkpv, Comm. Math. Phys. {bf 219} (2001) 199, [arXiv:hep-th/0006156]; L. D. Faddeev and R. M. Kashaev, J. Phys. {bf A35}
(2002) 4043, [arXiv:hep-th/0201049].}

\singlespace
\pagenumber=0
\normalspace
\medskip
\bigskip
\titlestyle{\bf{Black Hole Entropy and the Pseudo Goldstone Bosons of Conformal Symmetry}}
\smallskip
\author{ Edi Halyo{\footnote*{email: halyo@stanford.edu}}}
\smallskip
\centerline {Department of Physics} 
\centerline{Stanford University} 
\centerline {Stanford, CA 94305}
\smallskip
\vskip 2 cm
\titlestyle{\bf Abstract}
The very near horizon regions of nonextremal black holes have a conformal symmetry which is anomalous and spontaneously broken by the Rindler vacuum. Therefore, these black holes can effectively be described by the pseudo Goldstone bosons of conformal symmetry and their Schwarzian action which is determined by the breakdown of the conformal symmetry to $SL(2)$. 
In Euclidean gravity, the Schwarzian action leads to the Wald entropy of nonextremal black holes. These results are consistent with the neutral limit of near extremal charged black holes described by near $AdS_2$ space--times.

\singlespace
\vskip 0.5cm
\endpage
\normalspace

\centerline{\bf 1. Introduction}
\medskip

The are a number methods for computing the entropy of nonextremal black holes each with its own advantages and shortcomings. Unfortunately, none of these methods sheds light on the gravitational degrees of freedom that contribute to the entropy.
In this paper, we propose that these degrees of freedom are related to the pseudo Goldstone bosons (PGBs) of the conformal symmetry that appears in the very near horizon region. The conformal symmetry of the very near horizon region is broken down spontaneously and anomalously to $SL(2)$ giving rise to the PGBs. This symmetry breaking pattern completely determines
the Schwarzian action of the PGBs, which leads to the Wald entropy for nonextremal black holes.
Unfortunately, this does not fully resolve the question of the fundamental black hole degrees of freedom 
since it is an effective description that is based only on symmetry considerations.

It has been known for some time that the very near horizon region of any nonextremal black hole has a chiral conformal symmetry
[\CARL,\SOL]. This is simply the reparametrization symmetry of the dimensionless Euclidean Rindler time that arises on the horizon.
In refs. [\CFT], it was shown that a nonextremal black hole can be described by horizon conformal field theory (CFT) state with a central charge and conformal weight given by $c/12=L_0=E_R$ where $E_R$ is the dimensionless Rindler energy conjugate to the dimensionless Rindler time[\LEN]. Due to the exponential coordinate transformation between Euclidean and Rindler spaces the conformal weights in Rindler space are shifted by $-c/24$, i.e. $L_0^{\prime}=L_0-c/24$. The Cardy formula applied to this CFT state gives the correct Wald entropy for the black hole[\CFT].

The conformal symmetry that lives in the very near horizon region is broken spontaneously by the Rindler vacuum and explicitly by the conformal anomaly down to $SL(2)$. As a result, black hole physics at low energies is described the PGBs of the conformal symmetry
that parametrize $Diff(S^1)/SL(2)$.
The PGB action is simply the Schwarzian action which is fixed by the unbroken $SL(2)$ symmetry in the proper PGB fields[\VER]. We consider  the PGB action to be a gravitational one since the horizon CFT describes gravity in the very near horizon region. In Euclidean gravity, the Schwarzian action leads to the Wald entropy for nonextremal black holes. As a consistency check, we also consider the neutral limit of charged near extremal black holes which should correspond to Schwarzschild black holes. We show that the same PGB action is obtained in terms of the near $AdS_2$ boundary geometry. 

This paper is organized as follows. In the next section, we briefly review horizon CFTs as a description of nonextremal black holes. In section 3, we describe the same black holes in terms of the PGBs of the conformal symmetry that lives in the very near horizon region. In section 4, we consider the neutral limit of charged near extremal black holes in terms of their near $AdS_2$ geometries and show that they lead to the same PGB action. Section 5 contains a discussion of our results and our conclusions.

\bigskip
\centerline{\bf 2. Horizon CFT Description of Nonextremal Black Holes}
\medskip

In this section, we review horizon CFTs that describe nonextremal black holes.
It is well-known that the near horizon geometry of a nonextremal black hole in any theory of gravity is Rindler space.
Consider a black hole with a generic Euclidean metric of the form
$$ ds^2=f(r)~ dt^2+ f(r)^{-1} dr^2+ r^2 d \Omega^2_{D-2} \quad, \eqno(1)$$
in D dimensions. The horizon is at $r_h$ which is determined by
$f(r_h)=0$. If in addition, $f^{\prime}(r_h) \not =0$, the black hole is nonextremal and the near horizon geometry is described by Rindler space. Near the horizon, $r=r_h +y$ with $y<<r_h$, which leads to the near horizon metric
$$ds^2=f^{\prime}(r_h)y~ dt^2+(f^{\prime}(r_h)y)^{-1} dy^2+ r_h^2 d \Omega^2_{D-2} \quad. \eqno(2)$$
In terms of the proper radial distance, $\rho$, obtained from $d\rho=dy/\sqrt{f^{\prime}(r_h)y}$  and the dimensionless Euclidean Rindler time $\tau=(f^{\prime}(r_h)/2)~ t$ the near horizon metric becomes
$$ds^2=\rho^2 d \tau^2 + d \rho^2 + r_h^2 d \Omega^2_{D-2} \quad. \eqno(3)$$
This is the metric of Euclidean Rindler space times $S^{D-2}$ with fixed radius $r_h$. The Rindler metric in the $\tau$--$\rho$ directions naively looks like the flat metric in polar coordinates.  
The dimensionless Euclidean Rindler time in eq. (3) is an angle with a period of $2 \pi$. 

The dimensionless Rindler energy, $E_R$, conjugate to $\tau$ is obtained from the Poisson bracket[\LEN]
$$1=\{E_R,\tau\}=\left({\partial E_R \over \partial M}{\partial \tau \over \partial t}-{\partial E_R \over \partial t}
{\partial \tau \over \partial M} \right) \quad, \eqno(4)$$
where $M$ is the mass of the black hole conjugate to $t$. For large enough black holes, the rate of Hawking radiation is negligible and therefore we can assume that $E_R$ is time independent. Then, we find 
$$dE_R={2 \over f^{\prime}(r_h)}~ dM \quad.\eqno(5)$$
Using the definition of Hawking temperature, $T_H=f^{\prime}(r_h)/4 \pi$, eq. (5) can be written as
the First Law of Thermodynamics with the entropy given by $S=2 \pi E_R$. 
This procedure can be used for all nonextremal black objects with Rindler--like near horizon geometries in any theory of 
gravity[\SBH-\UNI].
In fact, it can be shown that $E_R$, which is a holographic quantity that can be obtained from a surface integral over the horizon[\EDIH], is exactly Wald's Noether charge $Q$[\WAL]. Therefore[\EDIW] the Wald entropy of nonextremal black holes is given by
$$S_{Wald}=2 \pi Q=2 \pi E_R \quad. \eqno(6)$$ 

In refs. [\CFT] it was shown that a nonextremal black hole with a near horizon geometry that is Rindler space can be described by a state of a 2D chiral CFT. The Rindler metric has a $U(1)$ symmetry that is simply the translation symmetry in the Euclidean 
Rindler time direction.
In the very near horizon region where $\rho \to 0$, the symmetry gets enhanced to a reparametrization symmetry of the Euclidean 
Rindler time, i.e. $\tau \to f(\tau)$. Reparametrization symmetry of a circle is $Diff(S^1)$ which is equivalent to one copy of the Virasoro algebra.  Thus we find that, in the very near horizon region of Rindler space, there is a chiral CFT which arises from the reparametrization invariance of the Euclidean time direction.
Alternatively, the conformal symmetry arises due to the fact that in the very near horizon region, energies are extremely redshifted and therefore all dimensionless quantities are negligible giving rise to scale invariance. In the two dimensional Rindler space this generalizes to the conformal symmetry.

In this chiral horizon CFT[\CFT], we identify the conformal weight of the black hole state in Euclidean space, $L_0$, with the dimensionless Rindler energy, i.e. $L_0=E_R$ and 
demand that the dimensionless Rindler temperature $T_R=1/2\pi$ be equal to the dimensionless CFT temperature, $T_{CFT}$, defined by
$$T_{CFT}={1 \over \pi} \sqrt{{{6 L_0^{\prime}} \over c}} \quad, \eqno(7)$$
where $L_0^{\prime}$ is the conformal weight of the state in Rindler space.
Rindler space is obtained from the Euclidean plane by an exponential transformation, i.e. the Euclidean coordinates $z=X+iT$ are related to the Rindler ones $u=\xi+i \tau$ by
$$z=\kappa^{-1}exp(\kappa u) \quad, \eqno(8)$$
where $\kappa$ is the surface gravity and $\kappa=1$ in the dimensionless Rindler coordinates in eq. (3).
As a result, the eigenvalues of $L_0^{\prime}$ are shifted relative to those of $L_0$[\CFT] giving $L_0^{\prime}=L_0-c/24$.

The entropy of a CFT state (in Rindler space) is given by the Cardy formula[\CAR]
$$S=2\pi \sqrt{{{c L_0^{\prime}} \over 6}} \quad, \eqno(9)$$
which can also be written in terms of $T_{CFT}$ as
$$S={\pi^2 \over 3}cT_{CFT} \quad. \eqno(10)$$
Using $T_R=T_{CFT}=1/2 \pi$ we find that the central charge of the chiral CFT is $c=12E_R$ and thus 
$$2 L_0^{\prime}=L_0={c \over {12}}=E_R \quad. \eqno(11)$$
Plugging these above values into eq. (9) or (10) then gives the correct Wald entropy in eq. (6).
Therefore, we can identify the very near horizon region of a nonextremal black hole with a 2D chiral CFT state that satisfies eq. (11)[\CFT]. The Cardy formula simply counts the different ways a CFT state at level $L_0^{\prime}$ can be realized and thus counts the black hole entropy in the microcanonical ensemble.

Note that this is not a usual CFT since $T_{CFT}=1/ 2 \pi$ is constant. It is better to think of the horizon CFT as a CFT in the Hagedorn phase in which the energy per degree of freedom 
is constant, i.e. $E/S=1/4 \pi$. When we increase the energy of the CFT, the temperature remains constant and energy increases due to the increasing number of degrees of freedom (with fixed energy). This suggests that the horizon CFT is somehow related to the Hagedorn phase of string theory but the connection is not obvious.

The Cardy formula is only valid asymptotically for $L_0^{\prime}>>c$ whereas our CFT state has $L_0^{\prime}=c/24$. This problem is usually solved by invoking fractionation, i.e. by assuming that there are twisted sectors of the CFT[\FAT]. 
The dominant contribution to entropy comes from the most highly twisted sector with a twist of $E_R$. Due to the twist,
the central charge of the CFT and the conformal weight of its states are effectively rescaled to
$c=12$ and $L_0^{\prime}=E_R^2/2$ respectively. From eq. (7) we see that after fractionation $T_{CFT}=E_R/2 \pi>>1$ i.e. the CFT is at a high temperature. The Cardy formula can now be applied since $L_0^{\prime}>>c$, and leads to the correct black hole entropy. 

\endpage

\bigskip
\centerline{\bf 3. Black Hole Entropy and the Pseudo Goldstone Bosons} 
\centerline{\bf of Conformal Symmetry}
\medskip

In the previous section, we saw that nonextremal black hole entropy can be obtained by using the Cardy formula for the entropy of a chiral CFT state that describes the very near horizon region. In this section, we obtain the same entropy from the physics of the 
PGBs of the conformal symmetry.

The conformal symmetry of the very near horizon region is broken spontaneously by the Rindler vacuum and explicitly by the conformal anomaly. First, it is clear that the CFT state that corresponds to the Rindler vacuum is at a nonzero temperature (given by eq. (7)) and spontaneously breaks the conformal symmetry. In fact,
the Rindler vacuum spontaneously breaks the full conformal symmetry (with generators $L_n$, 
$n=0,\pm 1,\pm 2, \ldots$) down to $SL(2)$ (with generators $L_0, L_{\pm 1}$). We will show this at the end of this section once we obtain the PGB solution that describes the Rindler vacuum.

Second, as usual, the conformal symmetry is anomalous. The anomaly shows up in the commutators of the Virasoro generators
$$[L_n,L_m]=(n-m)L_{n+m}+{c \over {12}} n(n^2-1) \delta_{n+m,0} \quad, \eqno(12)$$
where the second term proportional to the central charge is the anomaly. Note that there is no anomaly for $L_0, L_{\pm 1}$.
Thus, the full conformal symmetry is explicitly broken down to the global conformal symmetry or $SL(2)$ by the anomaly. As a result of the spontaneous and explicit
breaking of the conformal symmetry, at low energies, physics can be described by its PGBs. Since the global conformal symmetry remains unbroken, we expect the PGB Lagrangian to be $SL(2)$ invariant. 

Above, we saw that the chiral conformal symmetry in the very near horizon region corresponds to the reparametrization symmetry of the Euclidean Rindler time, $\tau \to \xi(\tau)$.
In ref. [\VER], it was shown that the PGBs of conformal symmetry can be described exactly by these reparametrizations 
which realize the conformal transformations nonlinearly. Then, the energy--momentum tensor for $\xi(\tau)$ is given by
$$T(u)=L_0^{\prime} \xi^{\prime 2}+{c \over {12}} Sch(\xi,\tau) \quad, \eqno(13)$$
where the Schwarzian is 
$$Sch(\xi,\tau)=\left({\xi^{\prime \prime} \over {\xi^{\prime}}} \right)^{\prime}- {1 \over 2} {\xi^{\prime \prime 2} \over {\xi^{{\prime}2}}} \quad, \eqno(14)$$
and prime denotes a derivative with respect to $\tau$. Using eq. (11) for the nonextremal black hole the CFT state has 
we can write eq. (13) as
$$T(\tau)=E_R \left({1 \over 2}\xi^{\prime 2}+ Sch(\xi,\tau) \right) \quad. \eqno(15)$$
From this we can deduce the PGB Lagrangian[\POL]
$$I_{PGB}=-E_R \int_0^{\beta} d\tau \left ({1 \over 2}\xi^{\prime 2}+ Sch(\xi,\tau) \right) \quad, \eqno(16)$$
For the dimensionless Rindler time $\beta=2 \pi$. 
This action describes the quantum mechanics of the field $\xi$ as a function of the Euclidean Rindler time $\tau$. The first term in eq. (16) is the the kinetic term whereas the second one describes the nonlinear interactions of the PGBs arising from the explicit breaking of the conformal symmetry. 
The PGB action above can also be derived from the geometric action of the Virasoro group[\VIR].

The kinetic term in the action in eq. (16) is not invariant under $SL(2)$ transformations of $\xi(\tau)$. Therefore, $\xi(\tau)$ is not the proper PGB field. Reparametrizing the PGB field by $\eta=tan(\xi/2)$ the PGB action becomes[\MAL]
$$I_{PGB}=-E_R \int_0^{\beta} d\tau  Sch(\eta,\tau)  \quad, \eqno(17)$$
which is invariant under $SL(2)$ transformations of $\eta(\tau)$. 
Thus, we conclude that it is in fact $\eta(\tau)$ that parametrizes the PGBs, i.e. $Diff(S^1)/SL(2)$, rather than $\xi(\tau)$. 
The $SL(2)$ invariance of the PGB action is due to the precise values of $L_0^{\prime}$ and $c$ for the CFT black hole state given in eq. (11) that gives rise to the correct coefficients for the kinetic and Schwarzian terms. This constitutes a nontrivial consistency check on the CFT state and the PGB description of the CFT.
The two actions in eqs. (16) and (17) are completely equivalent and in the following we will use the action for $\xi$ given in eq. (16).

The solution of the equation of motion for $\xi(\tau)$[\MAL], 
$$\xi^{\prime \prime}+\left[{1 \over {\xi^{\prime}}}\left({{(\phi_b \xi^{\prime})^{\prime}} \over {\xi^{\prime}}} \right)^{\prime} \right]^{\prime}=0 \quad, \eqno(18)$$
up to $SL(2)$ transformations, is given by $\xi(\tau)=(2 \pi/\beta)\tau=\tau$ or $\eta(\tau)=tan(\tau/2)$. In order to compute the entropy  associated with this solution we consider the PGB action in eq. (16) to be a gravitational action since we assume that the horizon CFT effectively describes gravity in the very near horizon region. Moreover, on very general grounds, 2D gravity is a CFT since after imposing the gravitational constraints, the only diffeomorphisms left are conformal transformations.
Therefore, we assume that the action in eq. (16) (for general $\beta$) is gravitational and and we can use it in Euclidean gravity. 
This leads to the partition function
$$log Z=-I_{PGB}=2 \pi^2 {E_R \over \beta} \quad. \eqno(19)$$
Thus, the entropy of the PGBs is 
$$S_{PGB}= 4 \pi^2 {E_R \over \beta}=2 \pi E_R \quad, \eqno(20)$$ 
where we set $\beta=2 \pi$. 
We see that the entropy of the PGBs of conformal symmetry is precisely the Wald entropy for nonextremal black holes. 

Unfortunately, the PGBs are not the fundamental black hole degrees of freedom but only give an effective description of the black hole based on the symmetry breaking pattern in the very near horizon region. Nevertheless, for the black hole state of the horizon CFT, the PGB description gives the correct entropy. We note that, unlike the Cardy formula, the above computation of entropy does not require $L_0^{\prime}>>c$ or fractionation. In fact, the exact values in eq. (11) were crucial in obataining the correct entropy.

We now show that the Rindler vacuum described by the PGB solution $\xi=\tau$ is $SL(2)$ invariant. The $SL(2)$ charges for the theory described by the action in eq. (16) (in Lorentzian signature) were obtained in ref. [\MAL]. $Q^0$ is given by
$$Q^0=E_R \left[{\xi^{\prime \prime \prime} \over {\xi^{\prime 2}}}-{\xi^{\prime \prime 2} \over {\xi^{\prime 3}}}- \xi^{\prime}
\right] \quad. \eqno(21)$$
For our solution, we find $Q^0=-E_R$. In addition, $Q^{\pm}$ are proportional to the second and third derivatives of $\xi$
and therefore vanish for our solution[\MAL]. We see that the charge $Q^0$ does not vanish even though we assumed that the Rindler vacuum is invariant under this symmetry. Following ref. [\MAL], we can explain this by considering the black hole in the thermofield double state, i.e. an eternal black hole. As a result, in the near horizon region we get two Rindler spaces (wedges) in the thermofield double state. It is this thermofield double state that is $SL(2)$ invariant with vanishing $Q^0$. The two Rindler wedges have equal and opposite charges with $Q^0_1=-Q^0_2$ and therefore the charge of any given Rindler wedge can be nonzero and negative.

For our purposes, the important point to emphasize is the fact that $S=2 \pi E_R=-2 \pi Q^0$, where  $Q^0$ is the Noether charge of (Lorentzian) Rindler time translations on the horizon.
Thus, the entropy we computed is the Wald entropy of nonextremal black holes as in eq. (6). As mentioned in section 2, the dimensionless Rindler energy is identical to Wald's Noether charge, i.e. $Q^0=-E_R$ in our case. This was realized before in ref. [\EDIW] but it is gratifying that we can derive this equivalence from the physics of PGBs of conformal symmetry.

We can also consider the Hamiltonian associated with $\tau$ translations which for our solution is[\MAL]
$$H={1 \over {2E_R}} (-Q^+Q^-+(Q^0)^2)={E_R \over 2} \quad, \eqno(22)$$
which is precisely the energy of the Rindler vacuum $L_0^{\prime}$.


\bigskip
\centerline{\bf 4. The Neutral Limit of Charged Near Extremal Black Holes}
\medskip

In this section, we consider the neutral limit of charged near extremal black holes. In this limit, these become
Schwarzschild black holes and we expect them to be described by the PGB Schwarzian action. 
It has been shown that the near horizon limit of near extremal charged black hole geometries has a near $AdS_2$ ($NAdS_2$) component
[\MAL]. After compactifying over the transverse directions, i.e. $S^{D-2}$, the physics of $NAdS_2$
is given by the action $I=I_{ext}+I_{nonext}$ where the extremal and nonextremal parts are[\POL,\MAL] 
$$I_{ext}=-{\phi_0 \over {16 \pi G}} \int d^2x \sqrt{g}R+2 \int_{bndy} du K \quad, \eqno(23)$$
and 
$$I_{nonext}=-{1 \over {16 \pi G}} \int d^2x \sqrt{g} \phi (R+2)+2 \int_{bndy} du \phi_{b}K \quad. \eqno(24)$$
Above, $\phi$ is the dilaton which parametrizes the area of the transverse $S^{D-2}$ and its VEV, $\phi_0=A_h$, is area of the extremal black hole horizon. $\phi_b$ is the boundary value of the dilaton which can be assumed to be a constant.
The one--dimensional boundary of $AdS_2$ is parametrized by the Euclidean boundary time $u$.
The extremal action in eq. (23) is topological and equal to $4 \pi$. Thus, $I_{ext}$ leads to the extremal entropy
$$S_{ext}=-I_{ext}={\phi_0 \over {4 G}}={A_h \over {4 G}} \quad. \eqno(25)$$
In the neutral limit $\phi_0=A_h=0$ and the extremal action vanishes.

The nonextremal action in eq. (24) is the Jackiw--Teitelboim (JT) action[\JT] with the additional boundary term. The dilaton equations of motion impose $R=-2$ and therefore the first term in eq. (24) vanishes. Thus, the theory is described simply by the boundary action.
In ref. [\MAL], it was argued that in the $NAdS_2$ theory the conformal symmetry of the boundary theory is broken down to the global conformal group $SL(2)$ just like in our description of the very near horizon region of Rindler space.
It was shown that, when the bulk coordinates of $AdS_2$, $t(u),z(u)$ in the Euclidean Poincare metric are considered to be functions of the boundary time $u$, the boundary action can be written as
$$I_{bndy}=-{1 \over {8 \pi G}} \int_0^{\beta} du \phi_b Sch(t,u) \quad, \eqno(26)$$
where the Schwarzian is given by eq. (14).
Here the $AdS_2$ bulk time $t(u)$ represents the PGBs of the broken conformal symmetry and the $SL(2)$ invariance of the PGB action is manifest.
The action for the $AdS_2$ boundary in eq. (26), with the identification $E_R=\phi_b/8 \pi G$, is identical to the one we obtained for the very near horizon region of Rindler space in eqs. (17). We remind that $E_R=S/2 \pi=A_h/8 \pi G$ and therefore we identify $\phi_b=A_h$, i.e. the boundary value of the dilaton is the horizon area.

We found that the neutral limit of $NAdS_2$ theory which corresponds to a Schwarzschild black hole, has an effective boundary action which is identical to the PGB Schwarzian action we obtained. This result is a nontrivial consistency check on our method of entropy counting in terms of the PGBs.
It may be surprising to obtain the same action for the very near horizon region of Rindler space and the boundary of $AdS_2$. This is due to the fact that the Schwarzian action is determined solely by the symmetry breaking pattern which is identical for both cases, i.e. conformal symmetry broken down to $SL(2)$.

\endpage

\bigskip
\centerline{\bf 5. Conclusions and Discussion}
\medskip

In this paper, we described the entropy of nonextremal black holes in terms of the PGBs of the spontaneously and anomalously broken conformal symmetry that arises in the very near horizon region. Even though the same entropy can be obtained from the Cardy formula for the CFT state that describes the black hole, our new description has the benefit of identifying the black hole degrees of freedom as the PGBs together with their action. The Schwarzian PGB action is completely determined by symmetry breaking pattern, i.e. the conformal symmetry broken to $SL(2)$. Unfortunately, this is not a microscopic counting of entropy since the PGBs are only effective degrees of freedom. We also supported our results by considering the neutral limit of near extremal charged black holes which is described by the same Schwarzian action.

As we saw above, the horizon CFT has a constant temperature $T_{CFT}=1/2 \pi$. Thus, it seems that the horizon CFT is in the Hagedorn phase where energy per degree of freedom is fixed to be $E/S=1/4 \pi$. Black holes have been described in terms of long strings with rescaled tensions at their Hagedorn temperatures[\SBH-\UNI]. It would be interesting to investigate whether there is a connection between these long strings and horizon CFTs. In particular, the description of the PGBs of conformal symmetry in the long string description seems to be obscure.

Our work is crucially based on the fact that the near horizon region of nonextremal black holes is Rindler space. However, it was recently shown that the entropy of these black holes can also be obtained from an $AdS_2$ space that lives at (Weyl transformed) infinity[\ADS]. The Schwarzian PGB action that led to black hole entropy above also appears as the boundary term in the (near) $AdS_2$ action. In some sense this is not surprising since the Schwarzian action is fixed by the symmetry of the problem: near both the origin of Rindler space and the $AdS_2$ boundary the conformal symmetry is broken down to $SL(2)$. Nevertheless,
it would be interesting to further investigate if these two space--times are more directly related.

Black hole horizons are chaotic[\CHA]. Thus, if nonextremal black holes are effectively described by the PGBs of conformal symmetry with the Schwarzian action, then this description should also lead to chaos. In fact, it can be shown that, the Schwarzian action of the PGBs, precisely in the Rindler coordinates leads to chaotic behavior[\VER,\NEXT]. The source of chaos is the nonlinear PGB interactions described by the Schwarzian term in eq. (16). This is an important consistency check on our description in terms of PGBs.

We obtained the nonextremal black hole entropy from the physics of PGBs of conformal symmetry which results from the reparametrization symmetry of the dimensionless Rindler time. On the other hand, as we saw above, the same PGB Schwarzian action arises from the physics of the $AdS_2$ boundary. In the $AdS_2$ context the Schwarzian action describes the fluctuations of the boundary. It is tempting to suggest that the same is also true in our case, i.e. the PGB action describes the fluctuations of an infinitesimal circle in Euclidean Rindler space around the origin which can be assumed to be the stretched horizon. 

As mentioned above, our description of nonextremal black holes in terms of the PGBs of conformal symmetry does not clarify the nature of the black hole degrees of freedom since it is an effective description based on the symmetries. What we need is
a microscopic theory on $S^1$ (the Euclidean Rindler time direction) that reproduces the physics of Rindler space just like the SYK model does for $NAdS_2$[\SYK,\MAL] (with the problems related to the dual bulk description notwithstanding). 
The SYK model itself cannot represent Rindler space because it has a finite entropy for vanishing temperature[\MAL]. However, the quadratic SYK model, i.e. what is usually called the $q=2$ case, has an entropy that is proportional to $T$[\MAL] and may describe the microscopic theory of gravity in Rindler space. This is basically a free fermionic model on $S^1$ with a random mass matrix. Unfortunately, since this model is basically free it is not chaotic as expected from a description of black holes.
Moreover, it is not symmetric under time reversal which may be problematic.
If it is possible to modify the random matrix model to give rise to chaos without destroying its appealing features, this would be a candidate for the microscopic description of nonextremal black holes.

On the other hand, it
is known that nonextremal black holes can be described by Liouville theory that lives on the horizon, i.e. the horizon CFT we used above is Liouville theory which satisfies eq. (11)[\LIO]. The Schwarzian PGB action is an alternative effective description of Liouville theory.
A candidate microscopic theory for this Liouville theory (or the Schwarzian PGB action) is the parafermionic Y model on a rhombic lattice[\YMO]. This model flows in the infrared to Liouville theory and has the same properties. Thus, we expect the parafermionic Y model to provide the microscopic description for nonextremal black hole horizons in analogy with the SYK model that microscopically describes the near $AdS_2$ space--times.


\bigskip
\centerline{\bf Acknowledgments}

I would like to thank the Stanford Institute for Theoretical Physics for hospitality.

\vfill

\refout

\end
\bye